\documentclass{article}
\usepackage{multirow}
\usepackage{tabularx}
\usepackage{array}
\newcolumntype{C}[1]{>{\centering\arraybackslash}m{#1}}
\newcolumntype{Y}{>{\centering\arraybackslash}X}
\usepackage{microtype}
\newcommand{\rej}[1]{{\text{Rej}_{#1}}}

\usepackage{graphicx}
\usepackage{subcaption}
\usepackage{booktabs} 

\usepackage{hyperref}

\usepackage[preprint]{icml2026}

\usepackage{amsmath}
\usepackage{amssymb}
\usepackage{mathtools}
\usepackage{amsthm}

\usepackage[capitalize,noabbrev]{cleveref}

\theoremstyle{plain}

\theoremstyle{definition}

\theoremstyle{remark}

\usepackage[textsize=tiny]{todonotes}

\icmltitlerunning{\textsc{PanopTag}: Simultaneously Tagging All Jets in a Particle Collision Event}

\begin{document}

\twocolumn[
  \icmltitle{\textsc{PanopTag}: Simultaneously Tagging All Jets in a Particle Collision Event}



  \icmlsetsymbol{equal}{*}

\begin{icmlauthorlist}
\icmlauthor{Umar Sohail Qureshi}{phys,slac}
\icmlauthor{Brendon Bullard}{slac}
\icmlauthor{Ariel Schwartzman}{slac}
\end{icmlauthorlist}

\icmlaffiliation{phys}{Department of Physics, Stanford University, Stanford, California, 94305, USA}
\icmlaffiliation{slac}{SLAC National Accelerator Laboratory, Menlo Park, California, 94025, USA}

\icmlcorrespondingauthor{Umar Sohail Qureshi}{uqureshi@cern.ch}

  \icmlkeywords{Jet Physics, Machine Learning, ICML, Transformer, Jet Tagging, High Energy Phyiscs, Particle Physics, Physics Data Analysis}

  \vskip 0.3in
]



\printAffiliationsAndNotice{}  

\begin{abstract}
Jet tagging, identifying the origin of jets produced in particle collisions, is a critical classification task in high-energy physics. Despite the revolutionary impact of deep learning on jet tagging over the past decade, the paradigm has remained unchanged. In particular, jets are classified independently, one at a time. This single-jet approach ignores correlations, overlaps, and wider event context between jets. We introduce \textsc{PanopTag}, a new paradigm for jet tagging that departs from traditional single-jet tagging approaches. Rather than classifying jets independently, \textsc{PanopTag} simultaneously tags all jets by employing an encoder-decoder architecture that uses jet kinematics as queries to cross-attend to particle flow object embeddings. We evaluate \textsc{PanopTag} on heavy-flavor $(b/c)$-tagging and demonstrate remarkable performance improvements over state-of-the-art single-jet baselines that are only accessible by exploiting event-level
features and correlations between jets.
\end{abstract}

\section{Introduction}
Particle colliders, such as the Large Hadron Collider (LHC) at CERN, accelerate beams of protons to nearly the speed of light and bring them to collide tens of millions of times per second. Each collision event produces complex sprays of outgoing particles that traverse a layered detector system \cite{ATLAS:2008xda, CMS:2008xjf} containing $\mathcal{O}(100\text{M})$ (with expected upgrades to $\mathcal{O}\text{(1B)}$) individual sensor elements of different types which are used to reconstruct particle trajectories, momenta, energies, and identities. From this raw detector information, physicists infer what particles were produced in the collision and whether the event contains evidence for new phenomena or novel processes.

A central object in this reconstruction chain is a jet \cite{PhysRevLett.39.1436, Sapeta:2015gee}, a collimated spray of particles originating from quarks or gluons that decay and shower almost instantaneously after being produced. Experimentally, jets serve as reconstructed proxies for the underlying quarks and gluons, defined by applying a jet clustering algorithm to the measured final-state particles. Jets are ubiquitous in collider events and appear in essentially every analysis at the LHC. One of the first and most important questions that physicists ask about a jet is ``what produced it?''. For example, jets initiated by bottom $(b)$ or charm $(c)$ quarks can carry information about Higgs boson or top quark decays, while jets initiated by gluons or light quarks often constitute background processes. Determining the origin of each jet, dubbed jet tagging, is thus a key classification task that underpins both precision measurements and searches for new physics.

From a machine learning perspective, jet tagging is a challenging prediction problem. The initiating particle radiates, those emissions radiate again, and so on, producing a cascade of roughly $\mathcal{O}(100)$ final constituents. Both the radiation cascade and the finite resolution and inefficiencies of the detectors smear the original particle's distinctive features, making its identity difficult to recover. Naturally, a jet is represented as a variable-size set of its reconstructed constituent particles, called particle flow objects (PFOs). Early approaches compressed this set into a small number of physics-motivated summary observables that were then used to construct multivariate discriminants to classify jets. However, in the past decade, modern deep learning methods \cite{hepmllivingreview} that operate directly on low-level constituent PFOs have become standard \cite{nature_hepml}. These methods often represent jets as images \cite{Cogan:2014oua, deOliveira:2015xxd}, sequences \cite{deLima:2021fwm}, or graphs \cite{Thais:2022iok, Qu:2019gqs}, but the prevailing state-of-the-art is to treat them as unordered sets or ``particle clouds'' and map these clouds directly to labels \cite{Qu:2022mxj} .
\begin{figure*}
    \centering
    \includegraphics[width=\linewidth]{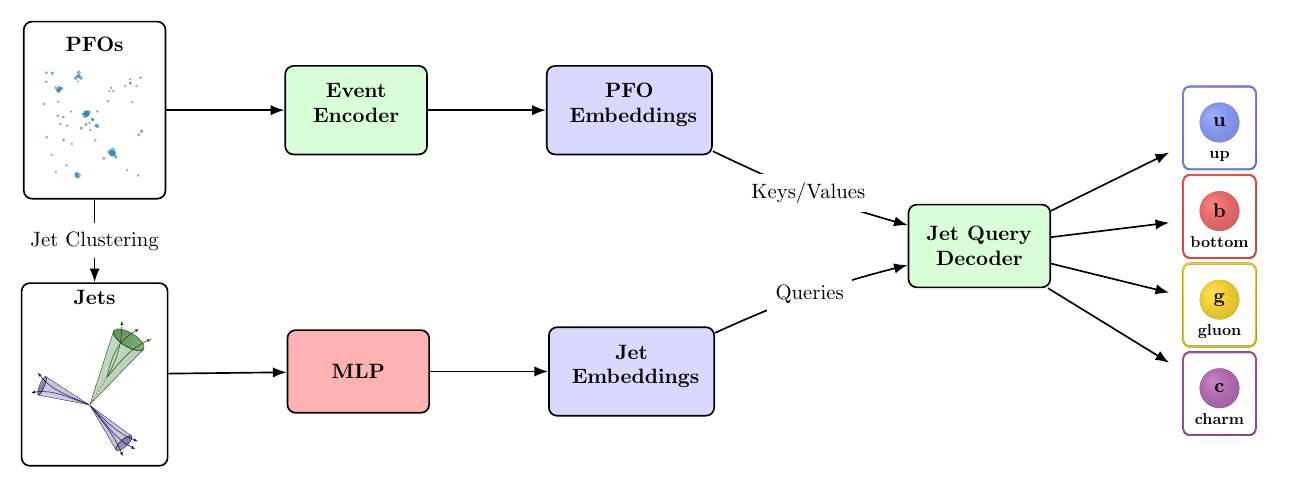}
    \caption{A high-level overview of the \textsc{PanopTag} architecture. PFOs are processed through the Event Encoder to produce embeddings, which serve as keys and values in a cross-attention mechanism. Simultaneously, jet kinematics are passed through an MLP to produce jet embeddings that act as queries, with the outputs after classification heads predicting class labels.}
    \label{fig:mainarchi}
\end{figure*}

In this work, we introduce for \textsc{PanopTag}, a new paradigm that reframes jet tagging as an event-level, multi-jet inference problem. Instead of classifying each jet and its associated constituents in isolation, \textsc{PanopTag} processes the full set of reconstructed jets and particles once with a DETR-inspired \cite{10.1007/978-3-030-58452-8_13} encoder–decoder architecture, assigning labels to all jets simultaneously. A permutation equivariant backbone first generates an embedding for all PFOs in the event. Jet kinematics are then used as queries in a decoder that cross-attends to the particle embeddings, producing one latent representation per jet, which is finally mapped to label probabilities as summarized in Fig.~\ref{fig:mainarchi}. This design allows information to be shared coherently across jets through a common particle representation, while letting each jet learn to focus, via attention, on the subset of particles most informative for its identity. We evaluate \textsc{PanopTag} on arguably the most critical jet tagging task, heavy-flavor identification, where it achieves remarkably higher performance than existing state-of-the-art single-jet baselines. We also demonstrate that the gain in performance is due to exploiting local correlations between nearby jets, which are not accounted for in traditional tagging algorithms.  

\section{Related Work}

Over the past two decades, considerable effort has been dedicated to developing deep learning methods for jet tagging \cite{Kogler:2018hem, Larkoski:2017jix}, largely aimed at shifting focus from physics-motivated hand-crafted features to learning directly from low-level jet constituents. Early successes explored multiple input representations e.g.~jet images, ordered sequences of particles, clustering trees, and particle graphs paired with convolutional, recurrent, and graph-based architectures. Recently, however, the dominant modality has been the particle cloud representation.

Within the particle-cloud paradigm, Particle Flow Networks (PFNs) and Energy Flow Networks (EFNs) \cite{Komiske:2018cqr} adapted the Deep Sets \cite{NIPS2017_f22e4747} formulation by treating a jet as an unordered set of constituents. Building on this idea, ParticleNet \cite{Qu:2019gqs} introduced a richer notion of geometry by dynamically constructing graphs and applying EdgeConv message passing from Dynamic Graph CNN (DGCNN) \cite{dgcnn}. More recently, attention-based approaches such as ABCNet \cite{Mikuni:2020wpr}, and the state-of-the-art Particle Transformer (ParT) \cite{Qu:2022mxj}, use transformers \cite{NIPS2017_3f5ee243} whose attention mechanisms are often augmented with pairwise kinematic information. In parallel, symmetry-aware designs \cite{Gong:2022lye, Bogatskiy:2022czk} have gained momentum, with models like LGATr \cite{spinner2025lorentz, Brehmer:2024yqw} incorporating Lorentz equivariance within the Geometric Algebra Transformer \cite{brehmer2023geometric} framework to improve robustness and generalization by explicitly respecting relativistic kinematics rather than relying on learned invariances.

Alongside these architectural developments, unsupervised \cite{Katel:2024ygn}, self-supervised \cite{Rieck:2025xnh}, and even fully-supervised methods \cite{Mikuni:2024qsr} including masked prediction \cite{Golling:2024abg, Leigh:2024ked}, contrastive objectives \cite{Dillon:2021gag}, and foundation-model-style pretraining on large datasets \cite{Bhimji:2025isp}, have been explored to leverage abundant simulation and real collision data for representation learning. Finally, as constituent multiplicities grow and event complexity increases with planned LHC upgrades, there has been sustained interest in efficient transformers \cite{Wang:2025dky} to reduce the quadratic scaling of full self-attention. 

It is worth noting that deep-learning-based jet tagging algorithms are now deeply integrated into LHC analyses, where they routinely improve object identification and thereby improve the precision of existing measurements and enhance the sensitivity in searches for new physics. Both CMS and ATLAS collaborations have developed and deployed a succession of high-performance taggers that take advantage of low-level constituent information to distinguish jets originating from $b$-quarks, $c$-quarks \cite{Mondal:2024nsa}, light quarks, and gluons, as well as jets originating from hadronic decays of top quarks \cite{ATLAS:2024rua} and electroweak $(W/Z)$ bosons \cite{Malara:2024zsj}. Recent efforts include transformer-based jet taggers from CMS \cite{CMS-DP-2024-066} and ATLAS \cite{ATLAS:2025dkv} collaborations. These developments underscore that continued progress in jet tagging is a key enabler for precision measurements and discovery at the energy frontier. 

\textsc{PanopTag} is complementary to the aforementioned research directions, but distinct at a more fundamental level. Rather than improving a single-jet classifier, it reframes jet tagging as an event-level, multi-jet inference problem. This formulation is able to exploit correlations and ambiguities that are difficult to express in the single-jet paradigm, e.g.~inter-jet kinematic constraints, overlap between nearby jets, and event-wide flavor structure. The primary obstacle to this approach is to ensure the classification generalizes to physics processes that are not included in the training, such that jet tagging efficiencies can be calibrated to data in control regions enriched in standard-candle processes (e.g. top quark pair production) and applied to arbitrary event topologies. In our study, \textsc{PanopTag} delivers a remarkable increase in performance over the ParticleNet and ParT baselines that generalizes beyond topologies included in the training dataset, demonstrating that event-wide context can unlock substantial performance gains that have remained inaccessible within the single-jet classification paradigm.

\section{Model Architecture}

\begin{figure}[h]
    \centering
    \includegraphics[width=\linewidth]{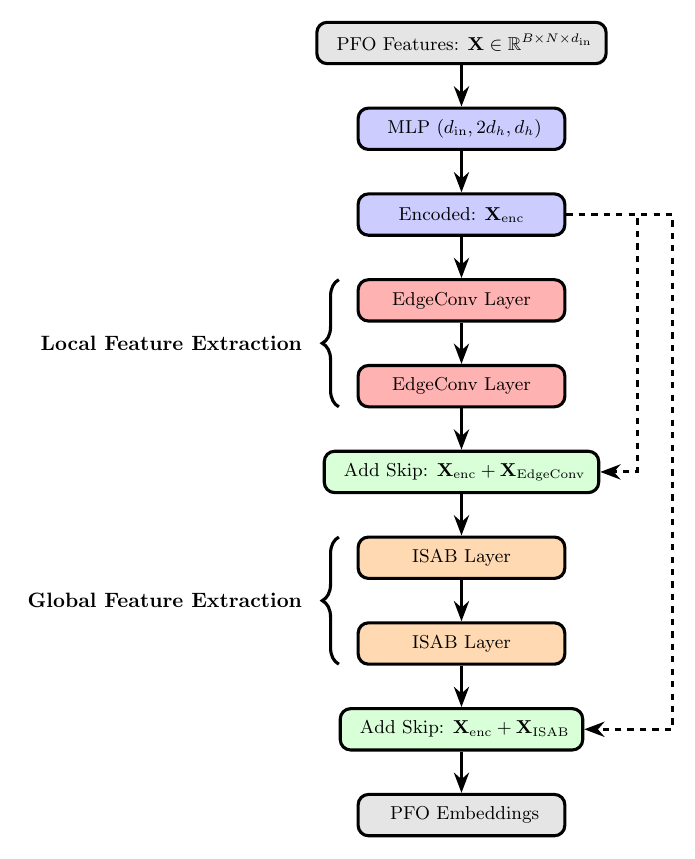}
    \caption{Illustration of the Event Encoder block. PFOs are first embedded using an MLP, then processed through two stages: (i) local feature extraction using EdgeConv layers for neighborhood aggregation, and (ii) global feature extraction using ISAB layers for set-based attention. Skip connections are added to preserve information flow across both stages.}
    \label{fig:encoder}
\end{figure}

The model consists of two main components, an Event Encoder (EE) block that processes all PFOs to compute contextual embeddings, and a Jet Query Decoder (JQD) block that cross-attends per-jet kinematic embeddings to these PFO embeddings to produce per-jet classification logits. 

\subsection{Event Encoder Block}
The EE combines local and global feature extraction to capture both short-range geometric structure and long-range event-wide context as summarized in Fig.~\ref{fig:encoder}. PFO input features $\mathbf{X} \in \mathbb{R}^{B \times N \times d_{\text{in}}}$ are first projected to the model dimension $d_h$ via an input MLP:
\begin{equation}
    \mathbf{X}_{\text{enc}} = \text{MLP}(\mathbf{X}).
\end{equation}
Local geometric structure is then captured through sequential EdgeConv \cite{dgcnn} layers. The first layer constructs $k$-nearest neighbor graphs in the PFO coordinate space (pseudorapidity $\eta$ and azimuthal angle $\phi$). For each PFO $i$, the $k$-nearest neighbors $\mathcal{N}_i$ are identified via Euclidean distance in coordinate space. EdgeConv computes edge features by concatenating relative differences with center points as: 
\begin{equation}
    \mathbf{e}_{ij} = [\mathbf{x}_j - \mathbf{x}_i , \mathbf{x}_i],
\end{equation}
for all $j \in \mathcal{N}_i$. These features are then passed through an MLP, and aggregated via max pooling, giving:
\begin{equation}
    \mathbf{x}^{\text{EdgeConv}}_i = \max_{j \in \mathcal{N}_i}\left(\text{MLP}(\mathbf{e}_{ij})\right).
\end{equation}
 Subsequent EdgeConv layers are then applied with Euclidean distances calculated based on the edge-convolved features. The local features after the EdgeConv layers are subsequently accumulated with skip connections. Our choice of this hierarchical design is motivated by \cite{pmlr-v251-kothapalli24a, Mikuni:2024qsr}, where incorporating local features is shown to generally improve performance. While transformers are capable of learning spatial interactions among particles, the addition of EdgeConv layers creates a latent representation that is better aware of the distances and topology. This is especially important for our event-level strategy since not only does \textsc{PanopTag} need to learn geometric correlations between particles to deduce their origins, but also to associate individual PFOs to jets. 

 \begin{table*}[t]
\centering
\caption{The set of per-particle input features used to train the models.}
\label{tab:infeats}

\begin{tabularx}{\linewidth}{@{} C{0.26\linewidth} C{0.10\linewidth} Y @{}}
\toprule
\textbf{Category} & \textbf{Variable} & \textbf{Definition} \\
\midrule

\multirow{4}{*}{Kinematics}
& $\eta$        & The pseudorapidity of the particle \\
& $\phi$          & The azimuthal angle of the particle \\
& $\log p_\mathrm{T}$  & The logarithm of the transverse momentum of the particle \\
& $\log E$             & The logarithm of the energy of the particle \\

\midrule
\multirow{6}{*}{Particle Identification}
& $q$      & The electric charge of the particle \\
& $e$      & Boolean flag, true if the particle is identified as an electron \\
& $\mu$    & Boolean flag, true if the particle is identified as a muon \\
& $\gamma$ & Boolean flag, true if the particle is identified as a photon \\
& $h^\pm$  & Boolean flag, true if the particle is identified as a charged hadron \\
& $h^0$    & Boolean flag, true if the particle is identified as a neutral hadron \\

\midrule
\multirow{4}{*}{Trajectory}
& $d_0$           & The transverse impact parameter of the track \\
& $d_z$           & The longitudinal impact parameter of the track \\
& $\sigma_{d_0}$  & The uncertainty in the transverse impact parameter \\
& $\sigma_{d_z}$  & The uncertainty in the longitudinal impact parameter \\
\bottomrule
\end{tabularx}

\end{table*}

 Following local feature extraction, global context is captured through Induced Self-Attention Blocks (ISABs) from Set Transformers \cite{pmlr-v97-lee19d}. Each ISAB performs two stages of attention by introducing learnable inducing points $\mathbf{I} \in \mathbb{R}^{ M \times d_h}$. First, the points attend to the full set of PFOs as:
 \begin{equation}
     \mathbf{H} = \text{MAB}(\mathbf{I}, \mathbf{X}^{\text{local}}).
 \end{equation}
 Subsequently, PFOs attend back to the compressed representation:
 \begin{equation}
     \mathbf{X}^{\text{ISAB}} = \text{MAB}(\mathbf{X}^{\text{local}}, \mathbf{H}),
 \end{equation}
  where $M$ is the number of inducing points. This approach reduces computational complexity from full self-attention's quadratic $\mathcal{O}(N^2)$ to linear $\mathcal{O}(MN)$ for fixed $M$. The resulting reduction in complexity is particularly important for our application since collision events can contain $\mathcal{O}(1000)$ PFOs, rendering full self-attention quite expensive.

  Each ISAB employs Multihead Attention Blocks (MABs), which implement the usual scaled dot-product attention, but without any ad-hoc positional encoding to ensure permutation equivariance:
  \begin{equation}
      \text{Attention}(\mathbf{Q}, \mathbf{K}, \mathbf{V}) = \text{SoftMax}\left(\frac{\mathbf{Q}\mathbf{K}^T}{\sqrt{d_h}}\right)\mathbf{V},
  \end{equation}
  split across $H$ heads, with residual connections and layer normalization \cite{ba2016layernormalization}. Global features are accumulated with a long-range skip connection, similar to local features. The final encoder output is masked to account for variable-length events, $\mathbf{H}_c = \mathbf{X}_{\text{final}} \odot \mathbf{X}_{\text{mask}}$, where $\mathbf{X}_{\text{mask}} \in \mathbb{R}^{B \times N}$ indicates valid PFOs. The \textsc{PanopTag} event encoder can equivalently be thought of as a hierarchical graph neural network with EdgeConv layers acting on a $k$-nearest neighbors graph and ISAB layers on a fully-connected graph, in which each node corresponds to a PFO.

  \subsection{Jet Query Decoder Block}

The JQD block generates per-jet predictions by cross-attending jet-specific query embeddings to the contextual PFO embeddings. Each jet is represented by its kinematic properties i.e.~its transverse momentum $p_\text{T}$, pseudorapidity $\eta$, azimuthal angle $\phi$, and mass $m$, which are embedded via an MLP to generate representations:
\begin{equation}
    \mathbf{Z}_0 = \text{MLP}_{\text{jet}}(\mathbf{J}),
\end{equation}
where $\mathbf{J} \in \mathbb{R}^{B \times M_{\text{jets}} \times 4}$ contains the kinematics of all jets in the batch. The decoder consists of $L_{\text{dec}}$ cross-attention layers, each computing $\mathbf{Z}_{l+1}, \mathbf{A}_{l+1} = \text{MAB}(\mathbf{Z}_l, \mathbf{H}_c)$, where jet embeddings are used as queries and PFO embeddings as keys and values. The cross-attention is made mask-aware to prevent attention to padded PFOs by masking attention scores before softmax:
\begin{equation}
    \widetilde{\mathbf{S}} = \mathbf{S} + (1 - \mathbf{M}) \cdot (-\infty).
\end{equation}
After the decoder layers, a final classification head projects jet embeddings to per-class logits as $\mathbf{L} = \text{MLP}(d_h, d_h, C)$, where $C$ denotes the number of classes. During training or analysis, attention weights from the final decoder layer can provide interpretability into which PFOs contribute most to each jet's classification decision.

\section{Heavy-Flavor Tagging Experiment}
Virtually all publicly available jet-tagging benchmarks, such as the top-tagging \cite{kasieczka_2019_2603256} and quark-gluon discrimination \cite{komiske_2019_3164691} datasets, as well as larger collections like \textsc{JetClass} \cite{JetClass}, provide data samples in a jet-centric format, where each example in is an isolated jet with no access to the full event context. Since \textsc{PanopTag} operates at the event level and requires all PFOs and jets within the same collision as input, we therefore simulate dedicated datasets. 

\subsection{Motivation}
Identification of jets originating from heavy-flavor quarks (such as bottom and charm quarks) at the LHC plays a critical role in its physics program. Improved performance in heavy-flavor tagging significantly enhances analyses that rely on $b$ and $c$-jets, including studies of Higgs boson pair production and measurements of the Higgs couplings to bottom and charm quarks. Both ATLAS and CMS have therefore made sustained efforts \cite{Mondal:2024nsa} to improve heavy-flavor tagging using increasingly powerful machine-learning taggers, thus making it an ideal application for \textsc{PanopTag}.

\begin{figure*}
\centering
    \begin{minipage}{0.495\linewidth}
    \centering
        \includegraphics[width=\linewidth]{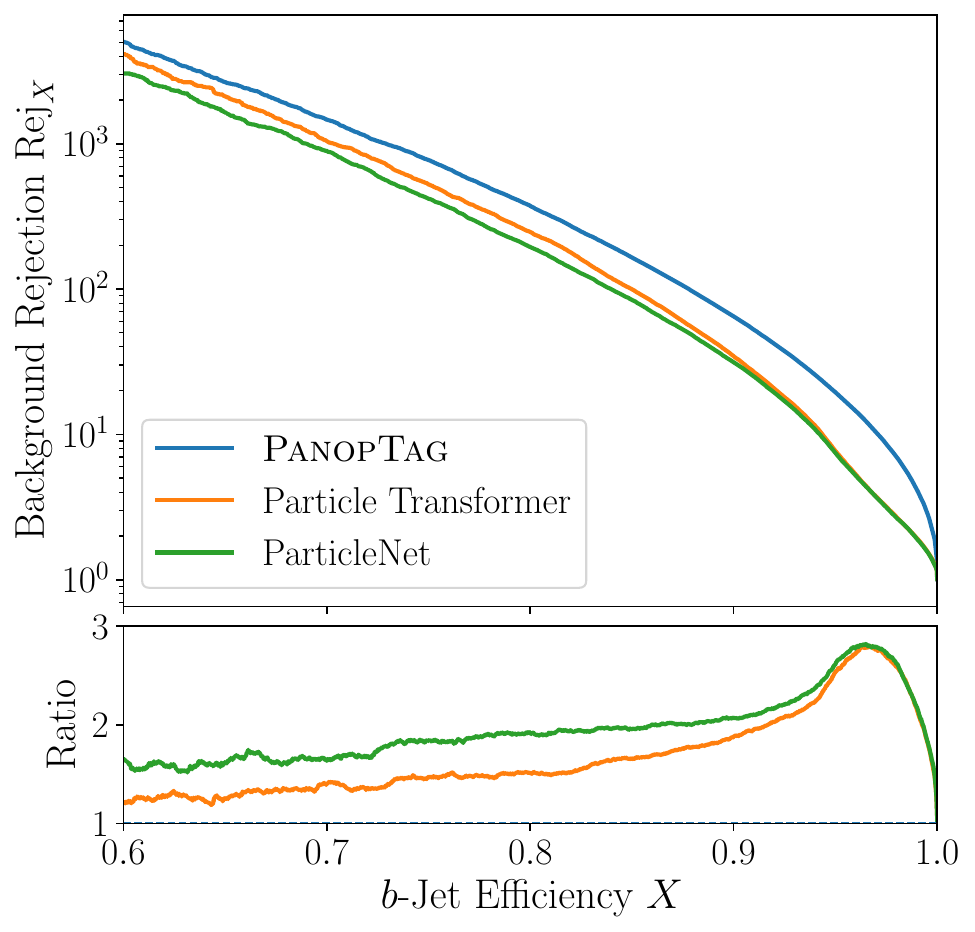}
    \end{minipage}
    \begin{minipage}{0.495\linewidth}
    \centering
        \includegraphics[width=\linewidth]{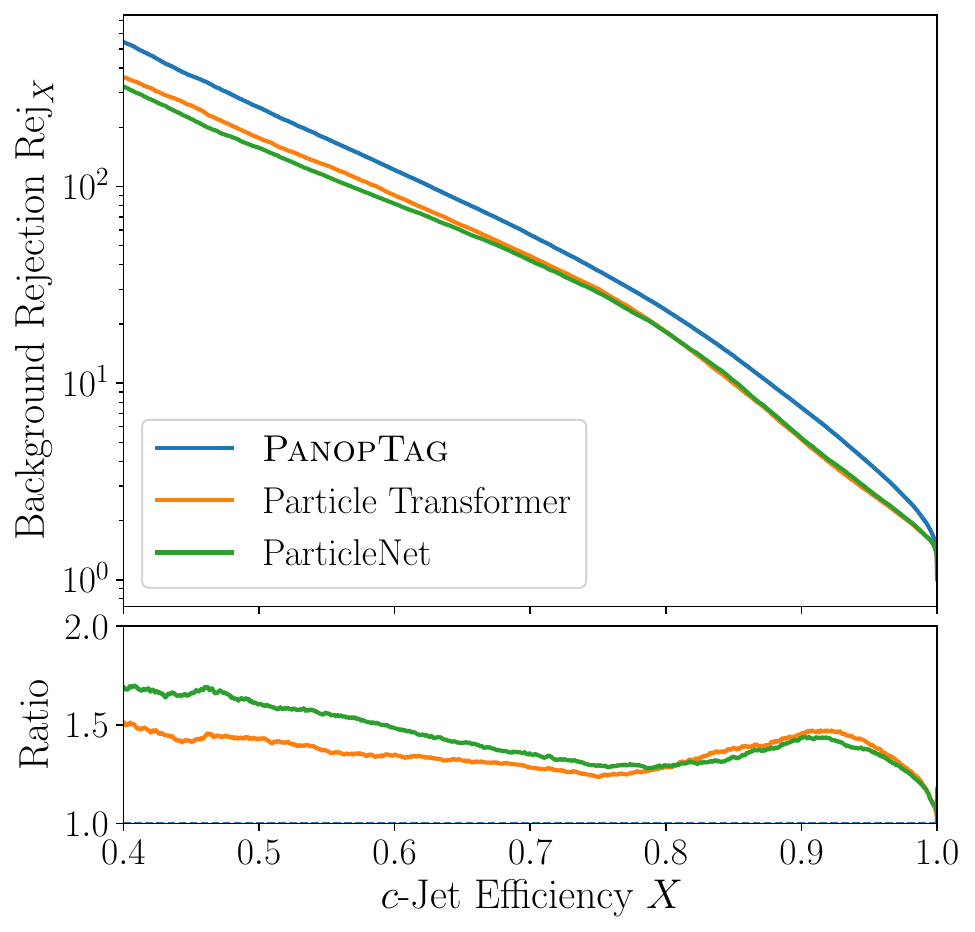}
    \end{minipage}
    \caption{The background rejections $\rej{X}$ as a function of the $b$-jet (left) and $c$-jet (right) tagging efficiencies $X$ for \textsc{PanopTag} (blue), ParT (orange), and ParticleNet (green) models. The ratio of the \textsc{PanopTag} background rejection to the baselines is shown in the bottom panel of each plot.\label{fig:sig_eff_bkg_rej}}
\end{figure*}

\subsection{Dataset}

Samples are simulated with standard Monte Carlo used by LHC experiments; in particular, we use \textsc{Pythia} \cite{Sjostrand:2014zea} for all physics event generation, in addition to account for parton showering and hadronization effects, yielding the final outgoing particles. To be close to realistic experimental conditions experienced at ATLAS and CMS, we use \textsc{Delphes} \cite{deFavereau:2013fsa} to simulate detector effects using the default ATLAS detector configuration provided in \textsc{Delphes}. Similar to existing datasets, we include multi-parton interactions but omit pileup interactions. Jets are subsequently clustered from reconstructed \textsc{Delphes} E-Flow PFOs via the anti-$k_\mathrm{T}$ algorithm \cite{Cacciari:2008gp} with a distance parameter $R=0.4$ and transverse momentum $p_\mathrm{T}>20$ GeV using \textsc{FastJet} \cite{Cacciari:2011ma, Roy_2023}. This setup is nearly identical\footnote{ATLAS and CMS employ more sophisticated, experiment-specific detector simulation and reconstruction software that is not available for studies performed outside the collaborations. \textsc{Delphes} is the de facto fast detector simulation used in most jet-tagging studies conducted externally, providing a widely adopted and reproducible approximation to detector effects for proof-of-concept algorithm development.} to the one used by ATLAS and CMS in their heavy-flavor tagging studies. Unlike the ATLAS and CMS studies, which consider only top–antitop ($t\overline{t}$) production, we also include single-top production, associated top-quark and $W$ boson production ($tW$), Higgs with associated $Z$ boson production ($ZH$), and top-antitop Higgs production ($t\overline{t}H$). This is done to ensure that \textsc{PanopTag} does not memorize the features of a particular topology, which would prevent its performance from extrapolating to processes not included in training. To evaluate this behavior, we train only on single-top, $t\overline{t}$, and $t\overline{t}H$, reserving $ZH$ and $tW$ for out-of-distribution validation in our testing. For each topology, we simulate 1M events, yielding a total of 5M events and about 30M jets in three classes i.e.~light (background) jets, $b$-jets, and $c$-jets, distributed as 55\%, 33\%, and 12\% respectively. The leading (descending order in $p_\mathrm{T}$) 128 PFOs are kept in each event, along with the 10 leading jets. For events with less than 128 PFOs and 10 jets, the entries are padded with zeros and ignored with masks during training and evaluation. For each PFO, we use the full gamut of information, including kinematics, particle identification, and trajectory displacement, as input features. The complete list of the 14 features for each particle is summarized in Table~\ref{tab:infeats}. The training of ParticleNet and Particle Transformer considers the same set of jets, but where associated PFO constituents are restricted to those that are used to form the jet at the clustering stage.

\begin{table*}[tb]
\centering
\begin{tabularx}{\textwidth}{@{} l *{6}{>{\centering\arraybackslash}X} @{}}
\toprule
& \multicolumn{2}{c}{\textbf{All Classes}} & \multicolumn{2}{c}{\textbf{$\boldsymbol{b}$-jets}} & \multicolumn{2}{c}{\textbf{$\boldsymbol{c}$-jets}} \\ \midrule
& Accuracy & AUC & $\rej{50\%}$ & $\rej{80\%}$ & $\rej{50\%}$ & $\rej{80\%}$ \\
\midrule
ParticleNet  & 0.910 & 0.9742 & 10,003  & 194 & 157 & 18  \\
ParT         & 0.913 & 0.9744 & 13,757  & 232  & 176  & 18  \\
\midrule
\textsc{PanopTag} &\textbf{0.935} & \textbf{0.9790} & \textbf{18,087}  & \textbf{383}  & \textbf{252} & \textbf{24} \\
\bottomrule
\end{tabularx}
\vspace{7.5pt}
\caption{Jet tagging performance on the heavy-flavor tagging dataset. For all the metrics, a higher value indicates better performance. The best-performing model for each metric is indicated in boldface. \label{tab:heavyflavorres}}
\end{table*}


\subsection{Setup}

A training-validation-testing split of 750k-100k-150k events is used for the single-top, $t\overline{t}$, $t\overline{t}H$ processes. For testing, we also include 150k $ZH$ and $tW$ events.  The base \textsc{PanopTag} model contains 2 EdgeConv layers and 3 ISAB layers in the event encoder block. The particle embedding dimension $d_h = 256$, the number of inducing points $m=48$, the number of attention heads $n_h = 32$, and the number of nodes for nearest neighbors graph construction $k=20$. All MLPs use the GELU \cite{hendrycks2023gaussianerrorlinearunits} activation function. The model in this formulation has 3.42M learnable parameters and fits comfortably on a single GPU. The PyTorch \cite{pytorch} library is used to implement the model.

The \textsc{AdamW} \cite{loshchilov2018decoupled} optimizer with a weight decay of $ 10^{-3}$ is employed to minimize the weighted focal cross-entropy \cite{Lin_2017_ICCV} loss function:
\begin{equation}
    \mathcal{L} = - w (1-p_\mathrm{t})^\gamma \log p_\mathrm{t}.
\end{equation}
Focal loss is chosen over vanilla cross-entropy because of its ability to dynamically focus on harder examples while down-weighting easy-to-classify jets. Class-weighting with weight $w$ is also applied during training to mitigate class imbalance. After mild hyperparameter tuning, the focus parameter $\gamma$ is set to 1. The learning rate is warmed up linearly over one epoch, starting at $10^{-6}$ and increasing to $2\times 10^{-4}$. We then apply cosine annealing \cite{loshchilov2017sgdr} with warm restarts every 15 epochs, and train for 30 epochs in total, using a batch size of 512. The performance is evaluated at the end of every epoch, and the checkpoint with the lowest validation set loss is used for all subsequent performance evaluation on the test set.

\subsection{Baselines}

We compare the performance of \textsc{PanopTag} to state-of-the-art single-jet baselines i.e.~the ParticleNet architecture adapted from DGCNN, and the ParT model, a transformer with a physics-motivated augmentation term added to its attention mechanism. All the models are trained end-to-end on the same dataset until convergence. For both ParticleNet and ParT, we use the existing PyTorch implementations from the \textsc{Weaver} \cite{weaver_core_2024} framework. The input features, optimizer, batch size, loss, and LR schedule remain the same as in the training of \textsc{PanopTag}, with the only key difference being that jets are inputted one-at-a-time. Moreover, instead of feeding the model absolute $(\eta, \phi)$ coordinates, coordinates $(\Delta\eta, \Delta\phi)$ relative to the PFO and the jet axis are inputted.  The peak learning rate is also re-tuned and chosen to be $5\times 10^{-3}$ for ParticleNet and $10^{-4}$ for ParT.

\subsection{Results}

For a thorough evaluation of performance, we adopt a series of metrics from both machine learning and physics jet tagging literature. Because jet tagging here is posed as a multi-class classification problem, we report two standard global metrics: the classification accuracy and the area under the ROC curve (AUC)\footnote{The AUC can be computed using the \texttt{roc\_auc\_score()} function in \textsc{scikit-learn} \cite{scikit-learn} with arguments \texttt{average=`weighted'} and \texttt{multi\_class=`ovr'}.}. In addition, we emphasize the background rejection metric, often used in high-energy physics. It is defined as the inverse of the false positive rate, evaluated at a fixed signal efficiency (true positive rate, TPR) of $X$\%: 
\begin{equation}
    \rej{X\%} \equiv 1/\text{FPR} \text{ at TPR}=X\%,    
\end{equation}
This quantity is reported separately for each signal $(b/c)$ jet category. By convention, light jets are treated as background, consistent with their treatment in LHC analyses, and each of the $b/c$ classes is considered as signal. The target signal efficiencies are chosen to reflect typical operating points at the LHC, and are set to 50\% and 80\%.

\begin{figure}
    \centering
    \includegraphics[width=\linewidth]{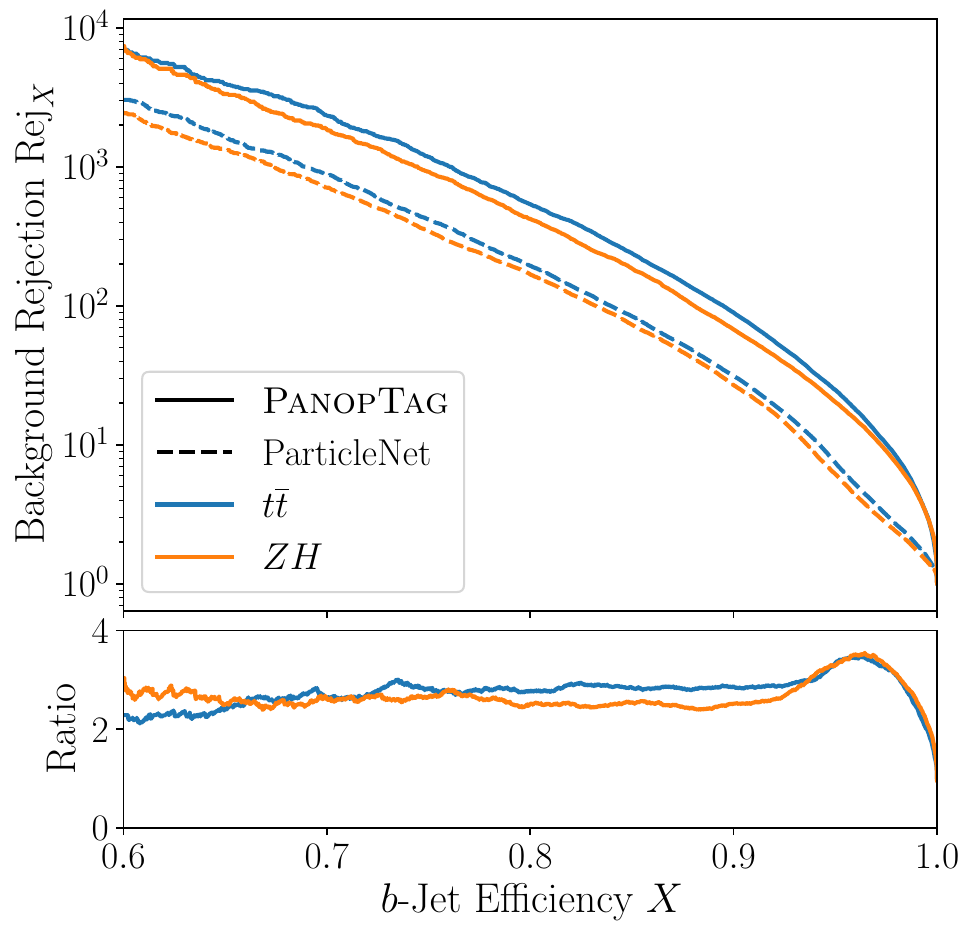}
    \caption{The background rejections $\rej{X}$ as a function of the $b$-jet efficiency comparing the \textsc{PanopTag} (solid) and ParticleNet (dashed) models for the $t\overline{t}$ (blue) and $ZH$ (orange) processes. The ratio of the \textsc{PanopTag} background rejection to ParticleNet is shown in the bottom panel for both processes.}
    \label{fig:OOD_plot}
\end{figure}

Although background rejection is uncommon in vision or language benchmarks, it is a standard merit in jet tagging because it directly maps onto discovery sensitivity (scaling as the square root of the rejection) at the LHC. For example, doubling the background rejection corresponds to an approximately 40\% gain in discovery potential, an improvement that would otherwise require roughly twice as much data, i.e.~about a factor-of-two increase in the LHC running time.

Table~\ref{tab:heavyflavorres} summarizes the performance of \textsc{PanopTag} and the single-jet baselines on heavy-flavor tagging. \textsc{PanopTag} delivers the best performance on every metric, and outperforms the existing state-of-the-art, ParT, by a wide margin. The overall accuracy on all classes is increased by 2.2\% and 2.5\% compared to ParT and ParticleNet, respectively. For the physics-oriented background rejection metric, \textsc{PanopTag} improves upon ParticleNet and ParT by a factor of 1.8 (2.0) and 1.7 (1.6) respectively for $b$-jets, and 1.6 (1.3) and 1.4 (1.3) for $c$-jets at a signal efficiency of 50\% (80\%). It is worth noting that the large improvement achieved by \textsc{PanopTag} is likely to lead to a significant increase in the discovery potential and sensitivity for new physics searches and precision measurements respectively at the LHC. In addition to reporting the numbers for background rejection at a fixed signal efficiency, Fig.~\ref{fig:sig_eff_bkg_rej} also shows the rejection metric as a function of $(b/c)$-jet efficiencies (TPR), where it is once again clear that \textsc{PanopTag} yields superior performance compared to existing models. More interestingly, the performance gains persist when evaluating on topologies not included during training, as illustrated in Fig.~\ref{fig:OOD_plot}, where \textsc{PanopTag} maintains its performance improvement on the $ZH$ process, which was not included in the training.

\begin{figure}
    \centering
    \includegraphics[width=\linewidth]{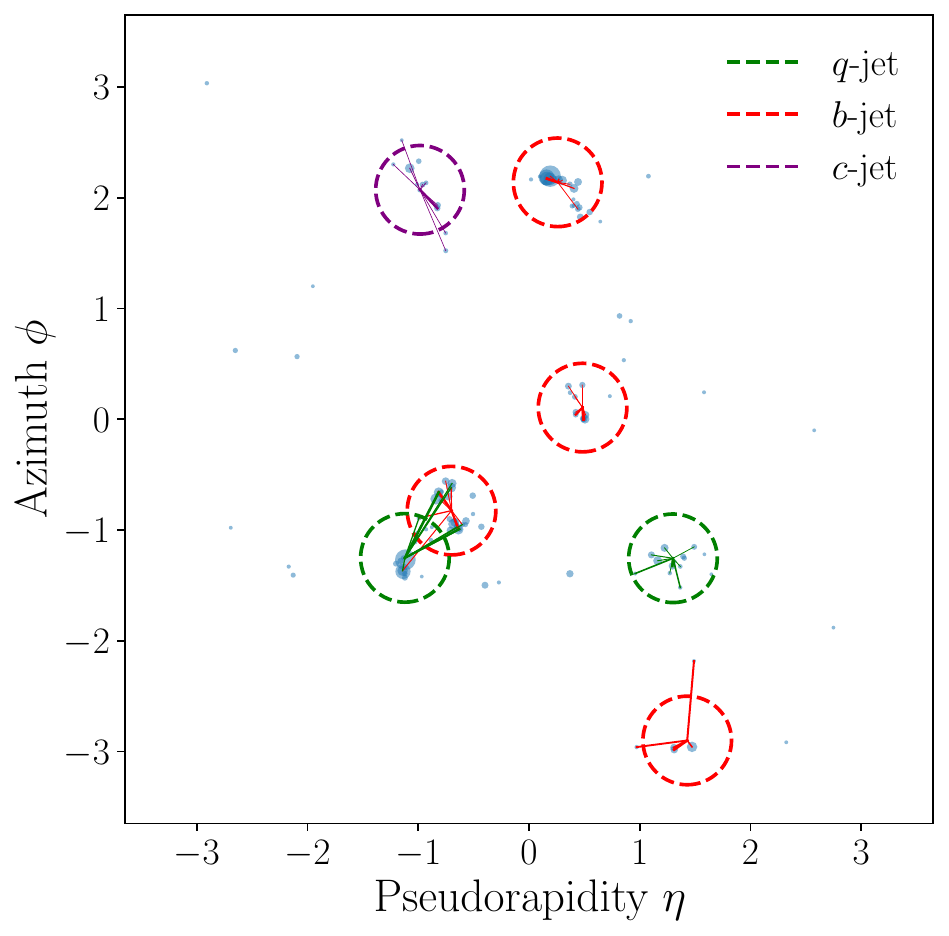}
    \caption{Representative event display in the $(\eta,\phi)$ plane. Markers show reconstructed PFOs with size proportional to $p_\mathrm{T}$. Dashed circles (of $\Delta R=0.4$) indicate jets, colored by flavor label. For each jet, we select the top-10 PFOs with the largest attention weights and draw line segments from the jet axis to these particles, visualizing the constituents most emphasized by the model. The width of line segments are proportional to the attention weights.}
    \label{fig:evdisp}
\end{figure}

\begin{figure}[h]
    \centering
    \includegraphics[width=\linewidth]{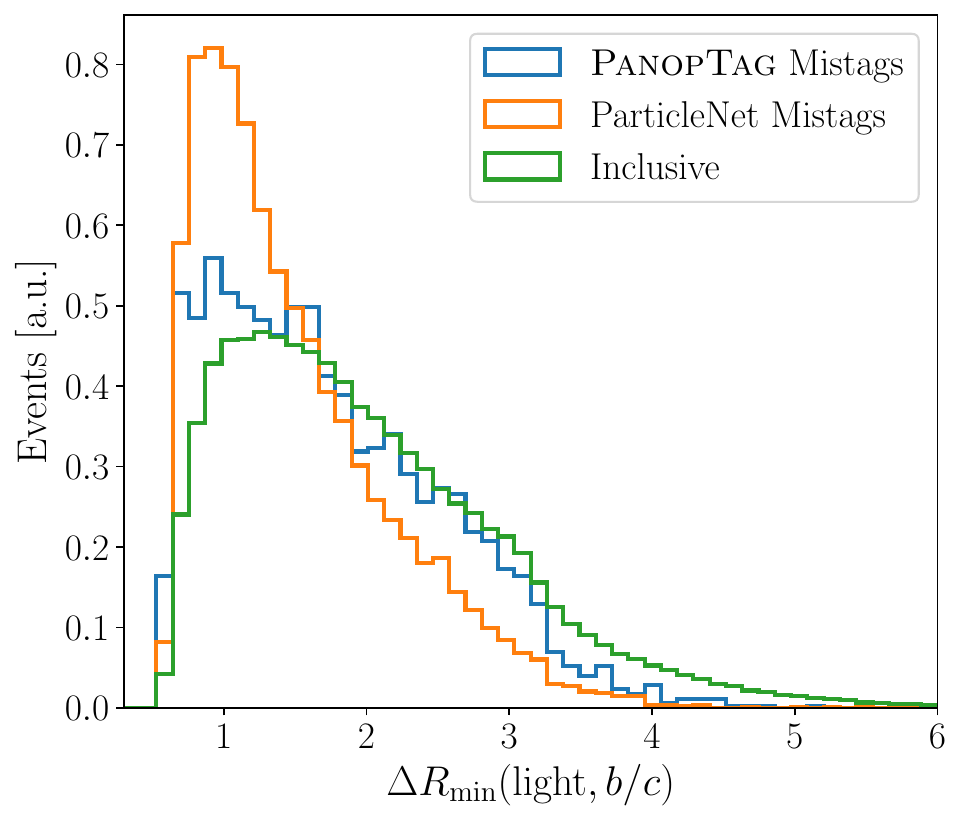}
    \caption{Angular distance $\Delta R$ between light jets and the nearest $b/c$ jet for jets mistagged by \textsc{PanopTag} (blue) and ParticleNet (orange), compared to the inclusive light jet distribution (green).}
    \label{fig:dR_to_heavy}
\end{figure}

\subsection{Opening the Black Box}
Interpretability is critical in high-energy physics to ensure that models learn physically meaningful correlations rather than dataset-specific artifacts. The final JQD layer provides a handle on interpretability, producing attention weights that can shed light on which PFOs are most relevant a jet's classification decision. Thus, the model's behavior can be checked against well-understood signatures (e.g.~locality) and tested for failure modes where the network might otherwise rely on spurious correlations. While a dedicated study similar to \cite{Mokhtar:2022pwm, Wang:2024rup, Legge:2025cnm} is required to fully understand the model behavior, we attempt to investigate where some of the performance gains originate.

Fig.~\ref{fig:evdisp} shows a visualization of the attention weights and the the particle cloud for a representative event in the $(\eta{-}\phi$ plane. Line segments from the jet axis extend to the top-10\footnote{The $b$-jet at $(1.42, -2.90)$ has only 4 constituents within the jet cone, so we limit the visualization to the top-5 most attented constituents.} most attended PFOs by the jet. The dominant feature is locality: for any jet, the highest-attended particles are overwhelmingly near the jet axis, consistent with expectation. More interestingly, the light jet at $( -1.12, -1.25)$, which is misidentified by both ParticleNet and ParT, strongly attends to particles in the $b$-jet near it.

Caution, however, is advisable before approaching these results. We only utilize attention weights from the final layer in the JQD and the JQD computes cross-attention between jets and PFOs subsequent to processing by the EE. This limits insights since earlier layers contribute to the representations; particles are convolved with each other, and one particle's embedding may be heavily influenced by and contain information about another. Additionally, prior work \cite{jain-wallace-2019-attention} has shown that attention weights should not be treated as a full explanation of a model's behavior.

Therefore, we also study the distribution of physical observables. Fig.~\ref{fig:dR_to_heavy} shows the $\Delta R \equiv \sqrt{\Delta \eta^2 + \Delta \phi ^2 }$ distance distribution of light jet to nearest heavy-flavor $b/c$ jet. Relative to the inclusive spectrum, ParticleNet mistags are strongly biased toward smaller $\Delta R$, indicating that ambiguities arise when a light jet lies close to a true $b/c$ jet and constituent overlap or wide-angle radiation makes the classification challenging. \textsc{PanopTag} shows a much smaller bias towards small $\Delta R$, following the inclusive distribution more closely, suggesting it can aid in suppressing proximity-driven mistags. Although this study highlights a key scenario in which context around jets provides additional information for classification in the presence of overlaps, a comprehensive investigation of the ways in which \textsc{PanopTag} outperforms the baselines is left to future work.

\section{Discussion and Conclusion}
Reconsidering how long-standing problems are posed is often a catalyst for innovation. In this work, we introduce \textsc{PanopTag}, a new paradigm which presents an event-level formulation of jet tagging. All jets in an event are simultaneously tagged by using each jet's kinematics as a learnable query that cross-attends to particle embeddings. Evaluated on heavy-flavor tagging, \textsc{PanopTag} yields significant performance gains over state-of-the-art single-jet baselines. We expect \textsc{PanopTag} to maintain or improve its relative performance with the inclusion of pileup effects, as these backgrounds produce a roughly uniform low-energy distribution of detector activity that is more readily identified using global event information. However, the training will require inclusion of a wide range of pileup conditions in order to generalize properly. While classification of jets is performed simultaneously in each collision event, the topology-agnostic performance enables the standard approach of single-jet tagging efficiency measurement and calibration to be performed in the usual fashion using $t\bar{t}$ control samples.


Beyond the numerical improvement, these results support a broader conclusion i.e.~treating jets as a set of correlated objects and enforcing a shared event representation is a productive modeling direction.  This is motivated by the physical reality that jets within an event are constrained by global energy-momentum conservation, characteristic production topologies, and even partial overlap or shared constituents.

Although we chose to focus on jet tagging, we stress that the \textsc{PanopTag} formulation is a more general recipe that can be repurposed for a wide class of jet problems. For instance, with small modifications to the final MLP head, \textsc{PanopTag} can support pileup-jet identification \cite{ATLAS:2017ywy}, per-jet underlying event corrections \cite{Haake:2018hqn}, jet $p_\mathrm{T}$/mass regression \cite{CMS-DP-2024-064}, background subtraction \cite{Qureshi:2025ylv}, and medium-response proxies in heavy-ion collisions \cite{Du:2020pmp}, etc. As such, the \textsc{PanopTag} architecture is a prime candidate for a high-energy physics foundation model.

This event-level perspective is also attractive from a fast-inference standpoint. Since the particle representation is computed once per event and reused across all jets, \textsc{PanopTag} avoids the repeated re-encoding implicit in running a single-jet tagger independently for every jet. In deployment regimes where latency and throughput are dominated by repeated forward passes and data movement, simultaneous tagging of all jets in a single forward pass provides a promising path to significant speed improvements, in addition to potentially increasing performance. Detailed studies of the inference speed of competing models is left to future work.

Finally, we emphasize that \textsc{PanopTag} should be viewed as complementary to ongoing research. The model is compatible with alternative backbones and e.g.~one can implement \textsc{PanopTag} with Lorentz-equivariant layers without changing the central premise of simultaneous tagging from a single event embedding. \textsc{PanopTag} is also a natural target for supervised, self-supervised, and unsupervised pretraining, since it exposes both particle and jet-level representations that can be trained with masking, denoising, alignment, or contrastive-style objectives, and subsequently adapted to downstream jet tasks with minimal changes.

In summary, these considerations suggest that event-level, multi-jet inference is a promising and extensible direction; it improves upon state-of-the-art performance today, while providing a template for extending collider machine learning toward unified, event-level solutions for a broad class of jet analyses tomorrow.

\section*{Acknowledgments}
The authors express their gratitude to Michael Kagan, Benjamin Nachman, and Caterina Vernieri for useful feedback on the manuscript. This work used the resources of the SLAC Shared Science Data Facility (S3DF) at SLAC National Accelerator Laboratory. SLAC is operated by Stanford University for the U.S. Department of Energy's Office of Science. This work is sponsored by the U.S. Department of Energy, Office of Science under Contract No. DE-AC02-76SF00515.

\section*{Data and Code Availability}
All code necessary to reproduce the results presented in this study is publicly available on GitHub at \url{https://github.com/umarsqureshi/PanopTag}.

\section*{Societal Impact Statement}
This paper presents work whose goal is to advance the field of machine learning in high-energy physics. There are many potential societal consequences of our work, none of which we feel must be specifically highlighted here.

\newpage

\bibliography{paper}
\bibliographystyle{icml2026}




\end{document}